\documentclass{elsart}
\usepackage{epsfig}

\journal{Nuclear Instruments and Methods}
\begin{document}
\newcommand{\cstwo}{CS$_2\,\,$  }
\begin{frontmatter}

\title{Negative Ion Drift and Diffusion in a TPC near 1 Bar}
\author{C. J. Martoff,} 
\author{R. Ayad,} 
\author{M. Katz-Hyman}
\address{Department of Physics, Temple University, Philadelphia, PA 19122, USA}
\author{G. Bonvicini,} 
\author{A. Schreiner}
\address{Department of Physics \& Astronomy, Wayne State University Detroit, MI 48202, USA}
\begin{abstract} 
Drift velocity and longitudinal diffusion measurements
are reported for a Negative Ion TPC (NITPC) operating with Helium + \cstwo 
gas mixtures at total pressures from 160 to 700 torr.  
Longitudinal diffusion at the thermal-limit was observed for drift fields 
up to at least 700 V/cm in all gas mixtures tested.  The results are
of particular interest in connection with mechanical simplification  
of Dark Matter searches such as DRIFT, and for high energy physics 
experiments in which a low-Z, low density, gaseous tracking detector 
with no appreciable Lorentz drift
is needed for operation in very high magnetic fields.  
\end{abstract}
\end{frontmatter}

\section{Introduction} A TPC which drifts negative ions 
(in this paper, CS$_2^-$) rather than electrons,
was invented to reduce diffusion in three dimensions
to its thermal (lower) limit
without applying a magnetic field\cite{negion,uclc,TPCSymp}.  
This provides the highest 3-D space-point resolution attainable for long
drifts, without the power requirements and expense of a magnet.

Such characteristics are particularly important for the development 
of the DRIFT series of 
direction-sensitive gaseous detectors searching for WIMP dark 
matter~\cite{drift}.
Three coordinates of good resolution on the recoil track are 
essential for DRIFT, in order to measure the length and direction
of tracks from low-energy atom recoils produced by elastic scattering
of massive WIMPs.
The standard solution of a TPC with magnetic field along the drift direction
would give good resolution in just two (transverse) 
coordinates. Furthermore the necessary large magnet is
impractical for underground experiments due to cost and electric power 
requirements.

Unlike the light electrons, negatively charged molecular ions 
are much more efficiently thermally coupled to the bulk of the gas
than drifting electrons would be.  In appropriate gas mixtures,
the negative ion drift mobility is near constant, and the rms 3-D
diffusion follows the ``low field" limiting behavior:
 
\begin{equation}
\sigma _D = \sqrt {\frac{ 4 \epsilon L}{eE}}
\label{drift}
\end{equation}

often up to reduced drift fields E/P of several tens of 
V/cm$\cdot$torr~\cite{tohru}. 

Of course
there are details to be reckoned with; the ions must form before the primary
ionization electrons drift far from their point of origin, or the resolution
will be spoiled from the beginning.  Also the negative ions must relinquish their
extra electron and produce a Townsend avalanche in the endcap gain
region.  Both of these requirements have been shown to be amply met by
\cstwo at 40 torr and by mixtures of \cstwo with small amounts of noble
gases at 40 torr total pressure\cite{tohru}.

The DRIFT I experiment\cite{drift} (active mass 0.16 kg) operates with 
a pure \cstwo fill at 40 torr , which is 
near-optimal pressure for a direction-sensitive WIMP search
with the DRIFT I spatial resolution\cite{cjmbuxton}.  

To achieve the much higher target masses 
planned for the DRIFT II (5 kg) and DRIFT III stages, higher target gas 
pressure and hence higher resolution are required.  However, the low 
recoil atom energy places an absolute upper limit on the target gas
density of less than 1 mg/cm$^3$. For pure fills of the medium-mass 
target gases that are most interesting as WIMP targets, this would require
running well below atmospheric pressure. The vacuum vessel and 
support then become a major element of cost and complexity, as they
are in DRIFT I. 

It is
therefore of great interest to see
whether a 1-bar gas mixture could be found which would 
not shorten the recoil tracks below any hope of directional detection,
but would still give all the benefits of negative ion drift.
In a previous paper, we reported on operation of GEM micropattern
gain elements in negative ion drift mixtures near 1 bar\cite{miyamoto}. 
Raising the total pressure using a Helium buffer gas is a natural 
solution to consider, since equal pressures of Helium and \cstwo have
densities in the ratio of approximately 4/76 = 0.05.  This report 
is to show that indeed mixtures of this kind do work well
as TPC gases. The 0.9 bar limit in the present work was imposed by the
apparatus and is not a limit of the technique itself. Such mixtures 
can therefore be considered for next-generation DRIFT detectors. 

Operation at 1 Bar also permits entirely new applications for NITPC. 
For example, a NITPC using a helium mixture has been
proposed for use as a main tracking detector in the 
NLC~\cite{uclc}.
The low drift speed of negative ions (only tens to hundreds of 
meters per second) allows arbitrary orientation of the drift direction 
relative to the momentum-measuring magnetic field, without 
producing any significant {\bf E $\times$ B} effects. The slow
pulse repetition rate and low duty factor of
NLC-like machines greatly mitigates the negative effects 
of the slow ion drift.

When combined with the very small diffusion broadening obtainable, 
the slow negative ion
drift velocity also brings phenomenal
z-resolution (along with good transverse resolution).  With sufficient gas 
gain and amplifier sensitivity, single electrons or clusters could be 
detected individually as in the TEC scheme, giving a number of 
statistical advantages 
for particle measurement\cite{walenta}.

\section{Experimental Methods}
\begin{figure}[h]
\label{Fig:mini_d}
\begin{center}\thicklines
\epsfig{file=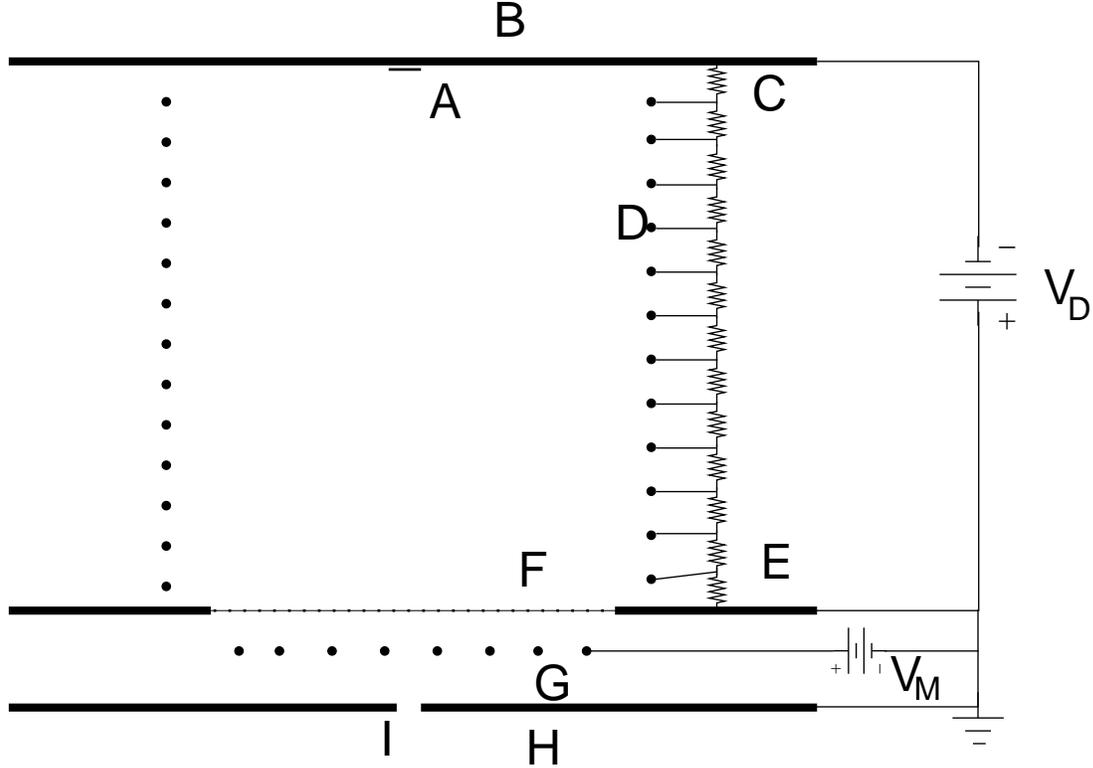,height=4.0in}
\caption{Schematic cross-sectional view of Mini-TPC.A: Sn photocathode, 
B: Drift cathode PCB,
C: Field cage voltage divider chain, D: Field cage, E Grid support
PCB, F: Grid mesh, G: Endcap MWPC anode wires, H: MWPC cathode PCB,
I: UV admittance aperture, V$_D$: Drift voltage, V$_M$: MWPC gain
voltage} 
\end{center}
\end{figure}

The tests were carried out with a small test TPC in a stainless steel bell 
jar with a simple gas manifold.  A sketch of the test TPC is shown in
Figure \ref{Fig:mini_d}.  The drift volume was rectangular, 50 x 60 mm 
transversely and 80 mm long in the drift (z) direction.  The field cage was
made of bare 500 micron diameter wires spaced 5 mm apart in z, 
stretched around nylon supports at the four corners. The drift-cathode was 
a solder-coated PCB with an Sn photocathode attached to it with conducting
epoxy.  

Charge was 
liberated from the photocathode by pulsed UV illumination from an EG\&G 
Flash-Pak~\cite{egg}.
The Flash-Pak's short-wavelength limit in air is $\sim$230 nm.  The 
standard internal capacitors of the Flash-Pak were augmented with 
additional HV capacitors to give a stored energy of about 0.2 Joule per 
pulse.  The Flash-Pak was triggered by an 
external pulser, from which a time-zero signal was also derived.
This system was a very convenient and cost-effective solution for 
generating variable-amplitude pulses of charge (photoelectrons)
in the TPC, which were sharply defined in time and space.  
It was found to be essential to scrape the photocathode clean
each time the detector was exposed to air between gas fills.

UV light entered the bell jar through 
collimating apertures and a quartz window, passed through a hole in the 
endcap cathode of the test-TPC, and struck the Sn photocathode.  
The endcap structure was an 8-wire MWPC.  One cathode of this MWPC (the 
``grid") terminated the drift-field region.  This grid 
was a transparent 
electrode made by epoxying a stainless steel mesh under tension, to a 
solder-coated  PCB which had a 50 x 50 mm window milled out of it.  The 
mesh\cite{wirecloth} pitch was nearly 40 cm$^{-1}$ and the geometrical
transparency was 81\%.  The anode plane of 8 Au-plated W wires
15 $\mu$m in diameter, at a pitch of 6 mm and with 50 g tension,
 was placed 6 mm behind the grid. The anode wires were  attached 
to a PCB anode frame using cyano-acrylate adhesive and soldered to 
contacts on the PCB.  The MWPC structure ended with a second cathode (the 
``MWPC cathode"), which was simply a solder-coated PCB.  i

Negative ``drift voltage"  up to -10 kV was
applied to the drift cathode; the grid and MWPC cathode were grounded, and 
positive high voltage was applied to the MWPC anodes.  This setup
has the 
obvious advantage that drift and gain voltages can easily be adjusted 
independently.  

Anode wires were read out individually through 1 nF high-voltage decoupling 
capacitors and amplified (and inverted) by Amptek A225 
amplifiers\cite{Amptek}.  The 
resulting positive signals were re-inverted by home-made common-emitter 
amplifiers, and then sent to the data acquisition system.

The gas system was based on a simple soft-soldered copper-tube 
manifold with attachments for 
introduction of various gases and of organic vapor (\cstwo~) from a small 
liquid reservoir.  Base pressure with the rotary pump used was 
about 50 milli-torr.  Negative ion drift chambers are very insensitive to 
contamination with anything less electronegative than \cstwo (for example, 
air).  It has 
been found that extraordinary precautions with gas purity are unnecessary.
Gas mixtures were prepared by admitting the minor component into the 
chamber first, followed by the major component.  Pressures were 
monitored with a mechanical gauge which is insensitive to 
the nature of the gas being measured.

Longitudinal diffusion was measured using the Amptek outputs and a digital
oscilloscope.  The NITPC output pulse width and its delay relative to the
FlaskPak trigger signal were measured as a function of
drift field, at constant MWPC voltage and constant
FlaskPak amplitude. The delay time was converted into drift speed using
the known geometry.  The amplifier shaping time was subtracted in quadrature
from the measured width.  The result was then converted from pulse width in time
to pulse width in distance by multiplying with the drift speed measured
at that drift field.  Space charge effects were shown to be absent by 
checking the results at high and low FlashPak amplitude.

\section{Results and Discussion}

Results for the four gas mixtures studied are shown in Table 
\ref{results}. To obtain the tabulated diffusion ``temperatures", the 
curves of FWHM diffusion width of the anode signals vs. 1/$\sqrt{E_D}$
were fitted to straight lines. The slopes were set equal to the
expected value for the FWHM of one coordinate in a 3-D diffusion problem,
obtained from Equation \ref{drift}:
\begin{equation}
FWHM = \frac{2.35}{\sqrt{3}} \sqrt{ \frac{6kTL}{eE_D}}
\label{eqn:class}
\end{equation}
and the corresponding temperature computed.

The deviations of the tabulated temperatures from the actual room temperature
of 293 K probably reflect the limitations of the simple diffusion
model used, rather than
any real physics.  Ohnuki et al\cite{tohru} using different detectors and
methods, also found diffusion temperatures differing significantly from
room temperature.
Typical data is shown in Figures 
3 and 4
for the case 200 torr \cstwo plus 500 torr He. A line fitted to the 
diffusion curve 
has a y intercept of -.01 rather than zero, and slope 1.30 rather than 1.49
expected from Equation~\ref{eqn:class}.

\begin{figure}
\begin{tabular}{|l|l|l|l|}
\hline
\label{results}
\cstwo (torr) & He (torr) & ion mobility & diffusion 
\\
&&${\rm {\Large\frac{cm/s}{ V/cm \cdot torr CS_2 }}}$ & ``temperature"\\
\hline
\hline
40 & 0 & 0.22 & 258 K\\
\hline
40 & 120 & 0.18 & 281 K \\
\hline
40 & 160 & 0.17 & 281 K \\
\hline
200 & 500 & 0.0071 & 229 K\\
\hline
\end{tabular}
\caption{Longitudinal diffusion results summary}
\end{figure}

\begin{figure}[h]
\label{Fig:vd}
\begin{center}\thicklines
\epsfig{file=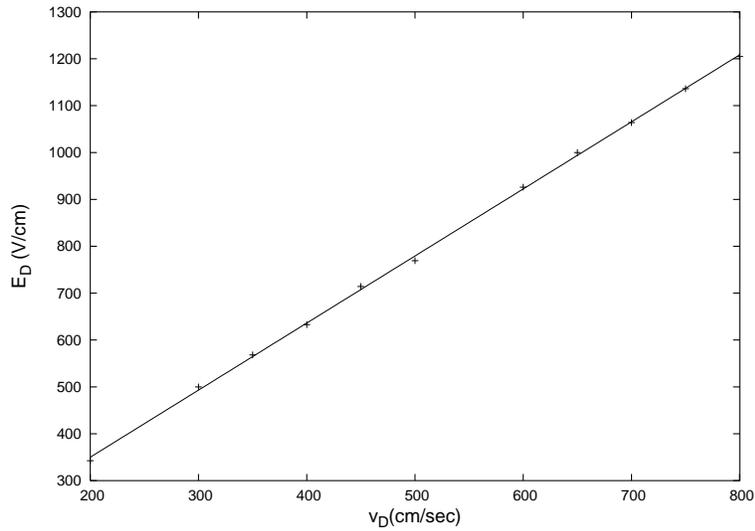,height=3.0in}=
\caption{Typical Drift Velocity Data. Measured drift velocity as
function of drift field for 500 torr He + 200 torr \cstwo.}
\end{center}
\end{figure}

\begin{figure}[h]
\label{Fig:diff}
\begin{center}\thicklines
\epsfig{file=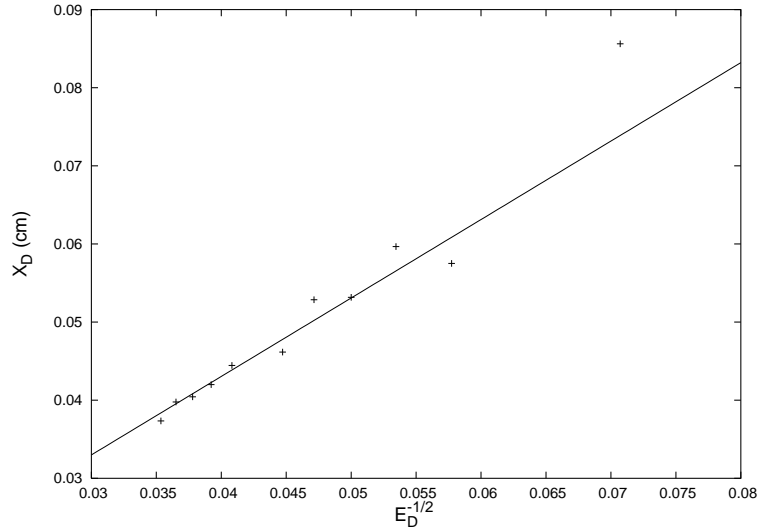,height=3.0in}
\caption{Typical Longitudinal Diffusion Data. Longitudinal diffusion
for 80 mm drift, measured as described in the text, for
500 torr He + 200 torr \cstwo.}
\end{center}
\end{figure}

For the lower total-pressure mixtures, the He acts essentially as a 
buffer gas, hardly changing the drift properties.  The drift 
mobility in the 700 torr mixture however is significantly lower than 
would be expected if it were dependent on the partial pressure of 
\cstwo alone.  The drift velocity itself drops 
by nearly a factor of six compared to the lower total-pressure mixtures.

\section{Conclusion}
Mixtures of \cstwo with a helium buffer gas were found to have
thermal-limit diffusion up to drift fields of over 12 V/cm$\cdot$torr.
These mixtures also allow stable 
operation of a Negative Ion TPC 
near 1 Bar total pressure.  This opens the way to numerous 
applications of the NITPC technique.  

\end{document}